
\magnification= \magstep1
\baselineskip = 24 true pt
\hsize = 16.5 true cm
\centerline{{\bf SPINNING PARTICLES  IN 2+1 DIMENSIONS }}
\vskip 1 true cm
\centerline{ Subir Ghosh}
\centerline{ Department of Physics}
\centerline{ Gobardanga Hindu College, 24 Pgs. (North)}
\centerline{ West Bengal, India}
\vskip 2 true cm
Abstract: Lagrangian and Hamiltonian formulations of a free
spinning particle in 2+1-dimensions or {\it anyon} are
established, following closely the analysis of Hanson and Regge.
Two viable (and inequivalent) Lagrangians are derived. It is also argued that
one of them is more favourable. In the
Hamiltonian
analysis non-triviaal Dirac Brackets of the fundamental variables are computed
for both the models. Important qualitative
differences with a recently proposed model for anyons are
pointed out.
\vfil
\eject

Particles in 2+1-dimensions are allowed to posess arbitrary spin
and statistics.  These so called anyons have been attracting a
great deal of attension [1] due to their varied applications in
planar physics, eg. fractional quantum Hall effect, high $T_c$
superconductivity and in processes in the presence of cosmic
strings. In our classical analysis as well, we will refer to the
spinning particles as anyons.

In order  to  proceed  to  the  field  theory  of  anyons,  it
becomes imperative to study the free anyon. An important step in
this  direction is the work by Jackiw and Nair [2], who have
derived an equation for the anyon, (in analogy to the Dirac
equation  for spin ${1\over 2}$ particle). Starting from a Poincare
group  representation  point  of  view,  they   have   taken   the
four-momentum  square  and  the Pauli -Lubanski scalar (since the
space  time  is  2+1-dimensional)  as  the   {\it   two   natural
independent}   choices  of  invariants.  This  is  important  for  our
formulation  as well since we will consider the model  which  has
the
above  characteristics,  as  the   more   favourable   one.On  the
other hand, a covariant Lagrangian description of anyon has been
attempted by Chaichian,Gonzalez Felipe and Martinez [3]. Also  in
[3],  a  Hamiltonian constraint  analysis  [4]  has  been  carried
through.  We have expressed our reservations regarding this model
at the end of the Letter.

In the present
Letter,  a  systematic  way  of  deriving  the  anyon  Lagrangian
has been
presented. This was formulated by Hanson and Regge  [5]  much
earlier,  for  the  case of spinning particles in 3+1-dimensions.
However, a detailed analysis reveals that there  are  interesting
differences  between  3+1 and 2+1-dimensional models, even at the
classical level. As we will elaborate, the formulation  is  based
    on deriving a consistency condition, involving the Lagrangian
$L$, the squared  linear  and  angular  momentum  and  the  basic
degrees  of  freedom  of  the  anyon. Allowed expressions for $L$
ought to come out  as  solutions  of  the  consistency  equation.
Unfortunately  this  equation  is too complicated to solve and an
eiconal  procedure  is  adopted  instead  [5]  in  deriving   one
particular  explicit  expression for an $L$. This model obeys the
famous Regge Trajectory (RT), that connects  its  mass  and  spin
value.  On  the  other hand, consistent with this scheme, another
type of $L$ is also possible which {\it does not  obey  the  RT}.
Indeed,  this  was  already  noted  in  [5]  under  some  special
circumstances. We argue that the latter model is more preferable,
since   it   also  agrees  with  the  basic  assumptions  of  the
Jackiw-Nair model [2]. Let us now move on to the actual analysis.

 Following Hanson and Regge [5], the anyon is depicted as a point
on the world line, to which a rotating  frame  is  attached.  The
(tentative)  Lagrangian  coordinates  are  taken  as the position
$x^{\mu}  (\tau)$,  and  angle  variables   $\Lambda   ^{\mu}_\nu
(\tau)$,        with       the       restriction
$${\Lambda
^{\mu}_{\phantom{a}\lambda}\Lambda^{\nu\lambda  }  =   \Lambda_\lambda
^{\phantom{a}\mu}
\Lambda  ^{\lambda  \nu}  =  g^{\mu \nu}; g^{00}=-g^{11}=-g^{22}=1
}\eqno(1) $$
where $\tau$ fixes the position on the  world  line.
The   velocities  are
$$\dot{x}^\mu  (\tau)={dx^{\mu}(\tau)\over
{d\tau}} \equiv u^{\mu}(\tau); \dot{\Lambda}^{\mu} _{\phantom{a}\nu}(\tau) =
{d\Lambda^{\mu} _{\phantom{a}\nu} (\tau )\over{d\tau}}. \eqno(2)$$
{}From (1)  we
find
$$\Lambda     _\lambda ^{\phantom{a}\mu}    \dot{\Lambda}^{\lambda
\nu}=\sigma^{\mu \nu} =-\sigma^{\nu \mu}. \eqno(3)$$
Thus $\sigma
^{\mu \nu}$ happens to be a better analogue of the usual  angular
velocity  (instead of $\dot{\Lambda}_\nu^{\phantom{a}\mu}$ ). Note the $\sigma
^{\mu \nu}$ contains more than the requisite number of components
since in two spatial dimensions, spin has  a  single  component.
Eventually  we  will  remove  the  extra  degrees  of  freedom by
imposing  constraints.  We  can  still   restrict   the   various
combinations  of  $u^\mu$  and  $\sigma^{\mu  \nu  }$ such that a
Poincare invariant Lagrangian  is obtained. This calls for a
subtle analysis [5] since {\it two} reference frames are involved
here, {\it i.e.} the (laboratory) fixed frame and the  one  fixed
on  the  anyon.  In the anyon frame $L$ will be a Lorentz scalar.
However, Poincare  invariance  demands  that  $L$  should  not  be
affected  by  translations  and rotations of the fixed frame. Thus
the variables comprising $L$ should be invariant (even though they  may
not  be Lorentz scalars), with respect to transformations of the
fixed  frame.  The  above  analysis  restricts   the   Lagrangian
variables  to only three scalar combinations $$a_1 = u^\mu u_\mu,
a_2    =    \sigma^{\mu    \nu}\sigma_{\mu    \nu},     a_3     =
u_\mu\sigma^{\mu\nu}\sigma_{\nu\lambda}u^\lambda.   \eqno(4)$$
A
further possible combination $a_4
=(\epsilon^{\mu\nu\lambda}u_\mu\sigma_{\nu\lambda} )^2$ depends
on  (4),
$$a_4=4a_3  + 2a_1a_2. \eqno(5)$$
The canonical momenta
are $$P^\mu=-{\partial L\over{\partial  u_\mu}}=  -2u^\mu  L_1  -
2\sigma^{\mu\alpha}\sigma_{\alpha \beta}u^\beta L_3, \eqno (6a)$$
$$       S^{\mu       \nu}=-{\partial       L\over      {\partial
\sigma_{\mu\nu}}}=-4\sigma^{\mu\nu}L_2   -    2(\sigma^{\nu\alpha
}u^\mu  u_\alpha  - \sigma^{\mu\alpha} u^\nu u_\alpha )L_3, \eqno
(6b)$$
where $L_i={\partial L\over{\partial a_i}}$. The equations
of motion are
$${\dot P}^\mu
=0, {\dot S}^{\mu\nu}   +    \sigma^{\mu\lambda}S_\lambda^{\phantom{a}\nu}    -
S^{\mu\lambda  }\sigma  _\lambda^{\phantom{a}\nu} =0 ={\dot S}^{\mu\nu} + u^\mu
P^\nu
-P^\mu u^\nu. \eqno (7)$$
The conserved Poincare generators  are
$P^\mu$  and  $M^{\mu\nu}=S^{\mu\nu}  +x^\mu P^\nu -P^\mu x^\nu$.
Note that in general the scalars
$$P^\mu P_\mu = M^2 = 4(a_1L_1^2
+2a_3L_1L_3   -{   1\over   2}    a_2a_3L_3^2),    \eqno    (8)$$
$$S^{\mu\nu}S_{\mu\nu}    =J^2    =8(2a_2L_2^2    +4a_3L_2L_3   -
a_1a_3L_3^2),         \eqno         (9)$$         $$W^2         =
(\epsilon^{\mu\nu\lambda}S_{\mu\nu}P_\lambda)^2={1\over 2} M^2J^2
-   S^{\mu\nu}P_\nu  S_{\mu\alpha}P^\alpha  ,  \eqno  (10)$$
are
operators having non-trivial Poisson Brackets (PB). (10)  is  the
square of the Pauli-Lubanski scalar.

     The next task is to specify the constraints such that only a
single component of the angular momentum $S^{\mu\nu}$ remains. An
appropriate   set   of   constraints   is  [6]
$$  V^\mu  \equiv
S^{\mu\nu}P_\nu \approx 0. \eqno(11)$$
(With this choice $W^2$ reduces to
the standard  expression.)  Note  that  only  two  components  of
$V^\mu$  are  independent  since  $V^\mu  P_\mu  =0$, and so they
reduce  $S^{\mu\nu}$  correctly  to  a  single  component,in  the
anyon frame..
Substituting  $P^\mu$  and  $S^{\mu\nu}$ from (6) in (11), we get
$$V^\mu  =   2\sigma   ^{\mu\nu}u_\nu   (4L_1L_2   -   2a_1L_1L_3
-2a_2L_2L_3  -2a_3L_3^2)  =0.  \eqno  (12)$$
For the validity of
(11), $L$ has to  satisfy  the  differential  equation  $$2L_1L_2
-a_1L_1L_3  -a_2L_2L_3  -a_3L_3^2  =0.$$  The choice of the above
condition  is  not  unique  and   indeed   other   (interrelated)
alternatives  are  available.  Actually  these  different choices
signify distinct frames of reference. The particular choice  (11)
specify  the  so called centre of momentum frame [7,5], where the
spatial momentum of the  system  vanishes.  The  centre  of  mass
coordinates   are   a   generalization  of  the  non-relativistic
definition, with the mass replaced by the  energy.  This  can  be
further  generalized  to  arbitrary inertial frames (see Pryce in
[7]).

Let us now introduce the  Lagrangian  $L$,  which  being  a
homogeneous function of degree one in the velocities, satisfies
$$u_\mu {\partial L\over {\partial {u}_\mu}} +{1\over 2}  \sigma  _{\mu\nu}
{\partial L\over {\partial \sigma _{\mu\nu}}} =L,$$ and is of the
generic form
$$L=2a_1L_1 +2a_2L_2 +4a_3L_3, \eqno (13)$$
obeying
the condition (12).

There are two approaches of deriving the desired Lagrangian.
The more fundamental and straightforward one
is  to solve a consistency condition involving $L$,  $M^2$,
$J^2$  and  $a_i$, obtained by eliminating the $L_i$ from (8),
(9),  (12)  and  (13). The relation  comes  in  fourth  power  of
$L^2$.  Each  root is a possible choice of $L$, producing its own
RT.
It has been shown in
[5] that for a relativistic spinning particle, in general $M^2$
and $J^2$ can not be fixed arbitrarily.  Rather
there exists an RT of the form
$$P^2 -f(S^2) =M^2 -f(J^2) =0.
\eqno (14)$$
In general the RT should
be obtained from  (8),  which relates $M^2$ to $J^2$, as in
(14). (However there is more subtlety involved  in  the  idea  of
having  a  non-trivial  RT for each $L$. But we will come back to
this point later.) What this means is that even classically,  the
mass   and   spin  of  a  particle can not be fixed indepentently
of each other, due to the
restriction imposed by the RT. For example, for a particular spin
value,  (remember  that  classically  the spin can be arbitrary),
only specific values of mass of the particle are allowed, which  are
computed  from  the RTs and each will have its own $L$, a root of
the consistency equation.  Clearly  all  these  models  are  {\it
inequivalent}.  However,  these  models  are somewhat unfamiliar,
since in general, their mass and spin are  operator  valued  (see
below  eqs.  (8)-(10)). Unfortunately the consistency equation is
very complicated to solve for $L$.

The alternative and more practicable approach is to proceed via the  RT
[5].  This  means that here one infers an explicit form of an RT,
(from other considerations, as for example a  linear  one  as  in
(16)  below),  and  then tries to construct an $L$, (by employing
the iconal procedure), that agrees with this RT and satisfies the
consistency relation as well, (such as the one in (15) below).
Generically the $L$ is reconstructed from the RT on
the boundary submanifold $a_3=0$ and  then  (12)  is used in a
generalized eiconal procedure [5] to continue $L$ for non-zero
$a_3$. An exampe of a non-trivial $L$ is
$$L^2 ={1\over 2}  [Aa_1
-Ba_2  +((Aa_1  -Ba_2)^2  -8ABa_3)^{1\over  2}],
\eqno (15)$$
corresponding to the given RT
$$BM^2 -{1\over 4} AJ^2 =AB. \eqno
(16)$$
(The derivation of this $L$ will be given in an enlarged
version.)

Let us now compute the canonical PBs between $x^\mu $,  $P^\mu
$, $\Lambda ^{\mu\nu} $ and $S^{\mu\nu}$. Since $\Lambda\Lambda
^T=g$ (1), there are only  three  independent  combinations  of
$\Lambda$ which we denote by $\phi_i, i=1,2,3$. These $\phi_i$'s
are  taken as   canonical   coordinates.   Let   us   also
define   $\sigma ^{\mu\nu}=a^{\mu\nu}_i(\phi)\dot{\phi}_i$.  The
conjugate  momenta  for $\phi_i$ are $$T_i =-{\partial L \over
{\partial\dot{\phi}_i}}  =  -{1\over 2}({\partial
\sigma^{\mu\nu} \over{ \partial\dot{\phi}_i}})({\partial L \over
{\partial\sigma^{\mu\nu}}}) = {1\over 2}a_i^{\mu\nu}(\phi)S_{\mu\nu}.
\eqno (17)$$
With these definitions a generic PB becomes
$$\eqalign{\{A,B\}&=({\partial
A\over {\partial x_\mu}})({\partial B\over {\partial P^\mu}} ) +
({\partial A\over {\partial \phi _i}})({\partial B\over {\partial
T_i}})
-({\partial A\over {\partial P^\mu _i}})({\partial B\over
{\partial x\mu}}) -({\partial A\over {\partial T_i}})({\partial
B\over{ \partial \phi _i}})\cr
&=({\partial A\over {\partial
x_\mu}})({\partial B\over{ \partial P^\mu}}) -({\partial A\over
{\partial P^\mu }})({\partial B\over {\partial x\mu}}) + S^{\mu
\nu}({\partial A\over{ \partial S^\mu _{\phantom{a}\lambda}}})({\partial B\over
{\partial S^{\nu\lambda}}})\cr
&+{1\over 2}(\Lambda^\mu_{\phantom{a}\alpha}
{\partial A\over {\partial \Lambda^{\mu\beta}}}
-\Lambda^\mu_{\phantom{a}\beta}{\partial A\over {\partial
\Lambda^{\mu\alpha}}}){\partial B\over{\partial S_{\alpha\beta}}}
-{1\over 2}(\Lambda^\mu_{\phantom{a}\alpha}{ \partial B\over {\partial
\Lambda^{\mu\beta}}} -\Lambda^\mu_{\phantom{a}\beta}{\partial B\over {\partial
\Lambda^{\mu\alpha}}}){\partial A\over{\partial S_{\alpha\beta}}}.}
\eqno(18)$$
With the help of (18), the PB's are
$$\{P^\mu,x^\nu
\}=-g^{\mu\nu}, \{P^\mu,P^\nu \}=\{x^\mu,x^\nu
\}=\{\Lambda^{\mu\nu},\Lambda^{\alpha\beta}\}=0 $$
$$\{\Lambda^{\mu\nu},S^{\alpha\beta}\}=\Lambda^{\mu\alpha}g^{\nu\beta}-\Lambda^{\mu\beta}g^{\nu\alpha},$$
$$\{S^{\mu\nu},S^{\alpha\beta}\}=S^{\mu\alpha}g^{\nu\beta}-S^{\mu\beta}g^{\nu\alpha}
+S^{\nu\beta}g^{\mu\alpha}-S^{\nu\alpha}g^{\mu\beta}.
\eqno(19)$$
The Hamiltonian description is pursued by first
noting that besides  the  RT,which is a First  Class  Constraint  (FCC),
(any  two) of the constraints$V^\mu$, in (12), constitute a
Second Class (SC) pair with $$\{V^\mu,V^\nu\}=S^{\mu\nu}M^2.
\eqno(20)$$
Consequently $\Lambda^{\mu\nu}$ should also be
constrained so that only one  angle variable remains. A
particularly useful covariant choice is [5]
$$\chi^\mu\equiv\Lambda^{0\mu}-{P^\mu\over M}\approx 0. \eqno(21)$$
Again
only two of the $\chi^\mu$ are independent since
$$\chi_\mu(\Lambda^{0\mu}+{P^\mu\over M})=0.$$ The   total
system  of  four  constraints  $V^1$,$V^2$,$\chi^1$  and
$\chi^2$ constitute a SC system and the RT (16) is the only FCC.
Thus the Hamiltonian is $$H=v(\tau)(P^2-f(S^2)), \eqno(22)$$
where $v(\tau)$ is an undetermined multiplier. It is
straightforward to  compute the  intermediate  Dirac  Brackets
(DB).  The  DB's  involving  $P^\mu$  remain unchanged, whereas
$x^\mu$ DB's are drastically altered to
$$\{x^\mu,x^\nu
\}_I=-{S^{\mu\nu}\over
{M^2}};\{x^\mu,S^{\nu\lambda}\}_I={{S^{\mu\nu}P^\lambda-S^{\mu\lambda}P^\nu}\over
{M^2}}.
\eqno(23)$$
Actually in the reduced system $S^{01}={S^{12}P_2\over P^0}$ and
$S^{02}=-{S^{12}P_1\over {P^0}}$, but the $\Lambda^{\mu\nu}$
components are not so simply related to, say$\Lambda^{12}$. That
is why we prefer to give the algebra in the covariant notation.
The final form is reproduced in (26).  The Poincare generators
$P^\mu$ and $M^{\mu\nu}$ DB's are
$$\{M^{\mu\nu},M^{\alpha\beta}\}_I=M^{\mu\alpha}g^{\nu\beta}
-M^{\mu\beta}g^{\nu\alpha}
+M^{\nu\beta}g^{\mu\alpha}-M^{\nu\alpha}g^{\mu\beta},$$
$$\{M^{\mu\nu},P^\alpha\}_I=g^{\mu\alpha}P^\nu
-g^{\nu\alpha}P^\mu .\eqno(24)$$
The single FCC, RT is still
remaining and $v$ in (24) is still arbitrary. But this is fixed
once a scale for $x^0$ is fixed, {\it eg.}
$$\Psi = x^0 -
t\approx 0. \eqno(25)$$
In conjunction with the RT (16), (25) is
a SCC system.So the final DB's are
$$\{x^\mu ,x^\nu \}_{II}=
-{S^{\mu\nu}\over {M^2}}+{P^\mu S^{0\nu}-P^\nu S^{0\mu}\over{M^2
P^0}};$$
$$\{P^\mu, P^\nu\}_{II}=\{P^\mu
,S^{\alpha\beta}\}_{II}=0; \{P^\mu,x^\nu \}_{II} =-g^{\mu\nu}
+{g^{\mu 0}P^\nu\over
{P^0}};$$
$$\{\Lambda^{\mu\nu},P^\alpha\}_{II}=({f'\over
{P^0}})g^{0\alpha}\Lambda^\mu_{\phantom{a}\lambda} S^{\lambda\nu};$$
$$\{x^\mu,S^{\nu\lambda}\}_{II}={S^{\mu\nu}P^\lambda -
S^{\mu\lambda}P^\nu\over {M^2}} -({S^{0\nu}P^\lambda
-S^{0\lambda}P^\nu\over {M^2 P^0}})P^\mu;$$
$$\{S^{\mu\nu},S^{\alpha\beta}\}_{II}=S^{\mu\alpha}(g^{\nu\beta}-{P^\nu
P^\beta\over {M^2}}) -S^{\mu\beta}(g^{\nu\alpha}-{P^\nu
P^\alpha\over {M^2}})$$
$$+S^{\nu\beta}(g^{\mu\alpha}-{{P^\mu
P^\alpha}\over {M^2}}) -S^{\nu\alpha}(g^{\mu\beta}-{{P^\mu
P^\beta}\over {M^2}}), \eqno(26)$$
where $f'={\partial
f\over{\partial J^2}}$, with $f$ obtained from (14). But here
 the $S^{\mu\nu}$ algebra is simple since
$$\{S^{12},S^{12}\}=0;\{x^i,S^{12}\}=-{{S^{12}P^i}\over {M^2}}.$$
All the $\Lambda^{\mu\nu}$ DB's are not shown, but it can be
seen,(see {\it eg.} $\{\Lambda^{\mu\nu},P^\alpha\}_{II}$ in (26)),
that only the {\it right} index of $\Lambda^{\mu\nu}$ transform
under Poincare generators. The Hamiltonian $$H=P^0=[P^2
+f(J^2)]^{1\over 2}, \eqno(27)$$ generates the time translations
$${dA\over dt}={\partial A\over\partial t}+\{H,A\}_{II}. $$ The final
Poincre DB algebra is the same as in (24).

However, due to the presence of nonzero DB between $x^\mu$'s and
the mixed DB's between position and angle variables, it is
better to switch over to the Pryce-Newton-Wigner variables
[5,7], in terms of which the DB's simplify and the variables are
amenable to quantization, such that the DB's are replaced by
commutators.

Now we shall discuss the special type of Lagrangian which was
mentioned before, which does not obey the RT restriction.
The following Lagrangian $${\cal L}^2 =[a_1 M^2 +{1\over 2}a_2 J^2
+2MJ({{a_1 a_2}\over 2} +a_3)^{1\over 2}], \eqno(28)$$
with $M$
and $J$ as {\it arbitrary numerical} parameters, also satisfies the
constraint relation (12) with $$P^\mu P_\mu =M^2, S^{\mu\nu}
S_{\mu\nu} =J^2, W=MJ. \eqno(29)$$ Thus in this particular case,
$P^2$ and $S^2$ can be numbers which are independently fixed.
(Note that in the 3+1-dimensional case in [5], this type of
solution appeared only in the limit of vanishing $a_4$, whereas
here since $a_4$ of (5) is not present from the very beginning, this is
also a solution without any approximation). If one were able to
solve explicitly the consistency condition mentioned before, (15)
and  (28)  should  appear  as  two of the independent roots. This is the
subtlety we mentioned before and for  reasons  obvious by  now,  we
believe (28) should be chosen to describe an anyon.

The constraint structure of this ${\cal L}$ in (28) is essentially
different since it has two independent FC's [2], $$\Phi_1 =P^\mu
P_\mu -M^2, \Phi_2 =\epsilon^{\mu\nu\lambda}S_{\mu\nu}P_\lambda
-MJ. \eqno(30)$$ (Due to the relation (10), $S^2$ is not
independent.) The $\Phi$'s are the two Casimir operators of the
2+1-dimensional Poincare algebra. The Hamiltonian is $$H=v_1
\Phi_1 +v_2 \Phi_2.$$ Therefore apart from fixing the time scale
to gauge fix the mass shell condition, (as in (15)), we have to
fix another gauge for $\Phi_2$, which we can take as $\chi^1$ of
(23). We still require two more constraints which are as in
(12). Interestingly the DB's due to the SCC's $\Phi_2$,
$\chi^1$, $V^1$ and $V^2$ remain the same as in (25).  The
reduced $H$ is $v_1 \Phi_1$. The final $H$ is$$H=P^0=[P^2
+M^2]^{1\over 2}$$.

Finally we come back to the model proposed in [3].
In order to forge a comparison between our results and those of
[3], the following identifications [8] are made, $$\sigma
^{\alpha\beta}\equiv {1\over 2}(n^\alpha\dot n^\beta -\dot n^\alpha
n^\beta), S^{\alpha\beta}\equiv  (n^\alpha p^\beta _n -p^\alpha _n
n^\beta), \eqno(31)$$ where $n^2=-1$ and $p^\mu_n=-{\partial
L\over \partial{\dot n^\mu}}$. Even though the DB structure of
[3] and ours agree, the Lagrangians are different. Our
expression in the new variables become $$\bar{{\cal L}}^2 =M^2{\dot x}^2
-{J^2{\dot n}^2\over 4} +MJ\epsilon^{\mu\nu\alpha}n_\mu\dot
n_\nu \dot x_\alpha, \eqno(32)$$ whereas in [3], the expression
is $${\tilde L}^2=m^2{\dot x}^2 -\alpha^2{\dot n}^2
+2m\alpha\epsilon^{\mu\nu\alpha}n_\mu\dot n_\nu \dot
x_\alpha+m^2({\dot x}^\mu n_\mu)^2. \eqno(33)$$
Notice that a term of the form of the last one in $\tilde L$ in
(33) is missing from $\bar{{\cal L}}$ in (32). This casts some doubt on
the validity of the Lagrangian in [3] because, (although naively
the Poincare algebra is satisfied), complications are involved in
the transformation properties. For example, as we mentioned
before, only the right index of $\Lambda ^{\mu\nu}$ transform
under Poincare generators, whereas no such distinction is
discernable in [3]. Furthermore, the constraint analysis of [3]
is very obscure. The authors have started with a simple
Lagrangian for the free anyon, where the spin parameter does not
appear in the Lagrangian. They then assume that the constraints
of this system will be preserved in the more detailed
Lagrangian, where the spin value explicitly appears. The adhoc
nature of introducing the Pauli-Lubanski constraint is
questionable. All these factors are probably responsible for the
mismatch between the Lagrangian in [3], (obtained in a heuristic
fashion), and our expression.

To conclude, we have derived two possible expressions for the
Lagrangian of a free anyon. One of them obeys the Regge
Trajectory restriction between its mass and spin values, whereas
the  other  does not. But in the latter model, the anyon mass and
spin are {\it independent c-number  parameters}.  Obviously  this
candidate  Lagrangian  is  a  better choice to describe an anyon,
instead of the one obeying RT and having operator valued mass and
spin. A detailed Hamiltonian analysis for both of these models is
carried through and
all the relevant Dirac Brackets are
computed. The variables suitable for quantization are also
discussed. Finally differences between our model and a
recently proposed model [3] are pointed out. The problems of
introducing interactions and a non-trivial gravitational
background are under study.

Acknowledgements: I thank Dr. P. Mitra and Narayan Banerjee for discussions.
\vfil
\eject
References:
\item{[1]} F.Wilczek, {\it Fractional Statistics and Anyon
Superconductivity}, (World Scientific, 1990)
\item{[2]} R.Jackiw and V.P.Nair, Phys.Rev.{\bf D43} (1991)1933
\item{[3]} M.Chaichian, R.Gonzalez Felipe and D.Lois Martinez,
Phys.Rev.Lett. {\bf 71} (1993)3405
\item{[4]} P.A.M.Dirac, {\it Lectures on Quantum Mechanics}, (Yeshiva
University Press, New York, 1964)
\item{[5]} A.J.Hanson and T. Regge, Ann.Phys.(N.Y.) {\bf 87} (1974)498;
see also A.J.Hanson, T. Regge and C. Teitelboim, {\it Constrained
Hamiltonian Systems}, (Roma, Academia  Nazi-  onale dei Lincei, 1976)
\item{[6]} J.Wessynhoff and A.Raabe, Acta Physica Polonica {\bf 9}
(1947)7; J.Wessynhoff, {\it ibid} (1947)26,34; J.L.Synge, {\it
Relativity: the Special Theory}, (North Holland, Amsterdam, 1955)
\item{[7]} M.H.L.Pryce, Proc.Roy.Soc. London, {\bf Ser A150} (1935)166;
{\it ibid} {\bf 195} (1948)62
\item{[8]} A. O. Barut, {\it Electrodynamics and the Classical Theory of
Fields and Particles},  (Mac-  millan Co., New York, 1964)
\vfil\eject\end